\pdfoutput=1

\documentclass{svjour2}                    
\smartqed  
\usepackage{graphicx}
\begin{document}

\journalname{Journal of Statistical Physics}

\title{Statistical Mechanics of Horizontal Gene Transfer in Evolutionary Ecology
}
\subtitle{}


\author{Nicholas Chia         \and
        Nigel Goldenfeld 
}

\institute{N. Chia \at
              Institute for Genomic Biology\\
              University of Illinois at Urbana-Champaign\\
              1206 West Gregory Drive, Urbana, IL 61801, USA \\
              Tel.: +001-217-3334425\\
              Fax: +001-217-2656800\\
              \email{chian@uiuc.edu}           
           \and
           Nigel Goldenfeld \at
           Department of Physics and Institute for Genomic Biology\\
           University of Illinois at Urbana-Champaign\\
           Loomis Laboratory of Physics\\
           1110 West Green Street\\
           Urbana, IL 61801, USA\\
              \email{nigel@uiuc.edu}
}

\date{Received: date / Accepted: date}

\maketitle

\begin{abstract}
The biological world, especially its majority microbial component, is
strongly interacting and may be dominated by collective effects.  In
this review, we provide a brief introduction for statistical physicists
of the way in which living cells communicate genetically through
transferred genes, as well as the ways in which they can reorganize
their genomes in response to environmental pressure.  We discuss how
genome evolution can be thought of as related to the physical
phenomenon of annealing, and describe the sense in which genomes can be
said to exhibit an analogue of information entropy.  As a direct
application of these ideas, we analyze the variation with ocean depth
of transposons in marine microbial genomes, predicting trends that are
consistent with recent observations using metagenomic surveys.

\end{abstract}

\section{Introduction}\label{intro}

The advent of high throughput sequencing technologies and their
application to genomics and metagenomics provides an ever-growing
torrent of data that is providing fine-grained data about ecosystems
and is beginning to alter our views on biological
evolution\cite{handelsman2004metagenomics,delong2005microbial}.  In
particular, it is becoming clear that the biological world, especially
its majority microbial component, is strongly interacting and may be
dominated by collective effects. Such phenomena can arise of course
through cell-to-cell communication by signaling molecules\cite{BASS05},
but there is also a genetic mode of communication: genes can be
transferred between living cells not related by heredity, and
subsequently expressed.  Comparative genomics indicate the widespread
and frequent presence of this so-called \lq\lq horizontal gene
transfer" (HGT) between organisms, not only those that are
closely-related but also those that are distant taxa.  Metagenomic
surveys also quantify the unsuspected extent of the abundance of mobile
genetic elements (MGEs) such as plasmids, viruses, and transposons, to
a degree that exceeds organismal abundances by an order of magnitude in
many different ecosystems\cite{frost2005mobile,ROHW05}.

The basis of classic Neo-darwinist population genetics rests on the
notion of gradual differences arising (i.e., via point mutation) within
a population. Populations are defined by firm geological and species
barriers that constrain the flow of genetic information. By contrast,
mobile genetic elements (MGEs) such as
plasmids~\cite{sorensen2005studying},
viruses~\cite{prangishvili2006viruses}, and
transposons~\cite{mahillon1998insertion} are capable of producing large
genomic changes and can jump between (classically defined) populations
by crossing species barriers. In particular, horizontal gene
transfer---the transfer of genes between organisms that are not related
by heredity---creates novel gene combinations by shuffling genetic
material between organisms that share the same environment. Such gene
transfers can have an enormous impact on the process of biological
evolution~\cite{ANDE66,ANDE70,sonea1988bwl,syvanen1994hgt,gogarten2005horizontal,frost2005mobile}.

The growing realization of the widespread nature of horizontal gene
transfer (HGT) is bringing about a re-evaluation of some of the
fundamental notions in biology, such as the species
concept\cite{goldenfeld2007biology,GOLD10}. Drawing general conclusions
about the frequency and effects of HGT requires massive amounts of
genome data. Experimentally, advances in high-throughput sequencing
have allowed researchers to sequence environmental samples identifying
microbial taxa and gene compositions~\cite{konstantinidis2009comparative}.

The lessons from statistical physics can be fruitfully translated to
evolutionary biology. Methods from statistical physics often utilize an
abundance of data in order to draw out general characteristics and
conclusions about the central features that describe system behavior.
Statistical approaches have already been implemented for many
biological systems including metabolic
networks~\cite{jeong2000large,thomas1995dynamical} and
ecology~\cite{hubbell2001unified_b,dewar2008statistical}. Statistical
physics is potentially important for understanding the emergent
properties of dynamical biological systems, such as the characteristics
and key features of evolution with HGT. Collective phenomena such as
synchronization in fireflies~\cite{mirollo1990synchronization} or pattern
formation in predator-prey dynamics~\cite{butler2009robust} rely on
techniques for studying many individual trajectories in order to arrive
at a condensed, more principled, understanding of system behavior. HGT
can be thought of as an interaction that couples the many
potentially different genomes in a population of organisms. Indeed,
cellular evolution as a whole finds the language of statistical physics
to often be the most appropriate for describing how modern cells
evolved~\cite{woese1977concept,woese2002evolution}.

In the remainder of this article, we briefly introduce the biology of
horizontal gene transfer and describe some of the ways in which HGT
plays an important role in biological systems. We then describe how HGT
requires an extension of ``classical'' notions of evolution, and how
this is being accomplished by recent efforts to model the effect of HGT
on genetic evolution. In order to predict the outcomes of HGT, concepts
from statistical physics can be useful, for example in interpreting
ecological data.  In particular we consider a striking trend that has
emerged from the sampling of environmental DNA extracted from marine
microbes, namely the variation in density of a particular type of
mobile genetic element (transposons) with depth in the ocean. We argue
that an intuitive notion of genome entropy, describing the variable
regions of a genome (ie. not the core, conserved genes), leads to
trends very similar to those observed.

\section{The extent of horizontal gene transfer}

Horizontal gene transfer, or HGT, is a ubiquitous feature of genome
evolution~\cite{eisen2000horizontal}. For the purposes of this article,
we consider an event to be a HGT if it involves any transfer or
introduction of any genetic material that does not stem from cellular
replication. Evidence for HGT is wide-ranging. It occurs within and
between all domains of
life~\cite{hotopp2007wlg,gladyshev2008massive,keeling2008horizontal,monier2009horizontal,pace2008repeated}.
The range of HGT encompasses the entire scale of organismal complexity,
from viruses~\cite{hendrix2000origins} to multicellular
eukaryotes~\cite{keeling2008horizontal}. Time is apparently not a
barrier: HGT events ranging from ancient~\cite{woese2000aminoacyl} to
very recent~\cite{holden2004complete} have been reported. Indeed,
there appears to be no absolute barrier to HGT, and we conclude that it
is a generic feature of genome dynamics.

Microbial genomes also contain and interact with a
large variety of mobile genetic elements (MGEs) including
plasmids~\cite{sorensen2005studying},
viruses~\cite{prangishvili2006viruses}, and
transposons~\cite{mahillon1998insertion}.
Counts of horizontally
transferred genes identified from G+C content found them to make up
anywhere from 1.5 to 14.5\% of microbial
genomes~\cite{garcia2000horizontal}. Characterizations of HGT in
different gene families found that 34\% of all gene families were
identified as having undergone HGT at some point in their evolutionary
history~\cite{cohen2010inference}.

Examples of genes transferred between evolutionarily distinct microbes
include rhododopsins in marine bacteria and
archaea~\cite{de2003proteorhodopsin,frigaard2006proteorhodopsin}. These
photosynthesis genes have been linked not to a particular organism so
much as a particular environment---so-called \lq cosmopolitan
genes'~\cite{frigaard2006proteorhodopsin}. These genes appear to be
readily transferred and incorporated into many different microbes, and
appears to aid them in harvesting light. The idea of a
gene being adapted and more tenaciously linked to an environment than a
particular microbe departs radically from the more classical notions of
vertical decent with mutation. In such cases, we are taking implicitly a more
gene-centric than organism-centric view of evolution.

While examples such as the photosynthetic rhododopsins depict a fairly
harmless tale of environmentally adapted genes, the same evolutionary
mechanisms are also responsible for a medical disaster that has arisen
due to the lack of appreciation for the potential swiftness of evolution.
Antibiotic and drug resistance genes rapidly adapt to new hosts,
meaning that once a single population of pathogens becomes resistant to
treatment, potentially many more rapidly will acquire resistance soon
thereafter~\cite{salyers1997antibiotic,bennett2008plasmid}. Moreover,
once the genes for antibiotic or drug resistance develop in some
pathogen they seem to persist over long periods of time in some form.
Thus, while avoiding a particular antibiotic treatment for long periods
of time has resulted in the loss of resistance, very rapid re-emergence
of resistance occurs once the antibiotic is
reintroduced~\cite{salyers1997antibiotic}. Interestingly, HGT of
antibiotic resistance genes from antibiotic-producing bacteria does not
appear to have played the major role in the evolution of antibiotic
resistance in the clinical setting~\cite{aminov2007evolution}. Instead,
most antibiotic resistance genes appear to have originated and
diversified in other environmental bacteria. They were then
disseminated widely, and these underlying genes formed the basis for
the development of antibiotic resistance in pathogens and commensal
bacteria~\cite{aminov2007evolution}. Resistance that seems to have been
developed in commensal microbes has made its way back to more open
environments in soil and water~\cite{koike2007monitoring}.

Potential barriers to HGT appear to be
abundant~\cite{thomas2005mechanisms}. First, genes must be delivered
into a microbe. This already presents a number of barriers. Viruses and
plasmids have limited host ranges~\cite{elsas2002ecology}. Naturally
competent microbes such as Neisseria are capable of uptaking raw DNA
from their environment (a process known as transformation), but uptake
is limited to DNA containing certain sequence
motifs~\cite{lorenz1994bacterial,chen2004dna}. Genes which are not
favorable are rapidly deleted and lost from a
genome~\cite{snel2002genomes}. In order to be retained on longer
timescales, acquired genes must pass through a gamut of hurdles. In
order to benefit its host, the newly-introduced gene must somehow be
expressed. Depending on the gene's history, it may now be subject to a
new regulatory scheme. Even once expressed, the gene may not be
expressed with the right timing or in the right amounts to benefit its
new host. Moreover, the newly expressed product has not yet been
adapted to the host environment, which may lead to a greater chance of
unwanted or deleterious interactions with other proteins in the new
cellular environment. Ultimately, the horizontally-transferred gene
must bypass the many levels of regulation and positively impact the
organism it has entered in a short time in order to be retained.

These barriers appear to be more surmountable the more closely related
are the donor and recipient, and as such, HGT is generally believed to
occur more often between the same or similar
species~\cite{elsas2002ecology}. This tendency is amplified by the fact
that organisms of the same species are spatially-correlated also.
However, overall there appear to be no absolute barriers to HGT despite
the numerous barriers to any individual trial. Along with how many HGT
events are retained in the long term, another pertinent question to
raise is how frequently does HGT occur on a shorter timescale? By
directly counting the number of times each plasmid horizontally
transferred in a population of lab-grown \textit{Escherichia coli},
Babic \textit{et al.} were able to determine that plasmids were
transferred approximately once per cell
generation~\cite{babic2008direct}. Thus, while the barriers to
individual HGT events persisting in a foreign host for very long
might be low, the number of attempts are seemingly high.

\section{Detecting horizontal gene transfer}\label{background}

Despite the extent of HGT, the evolutionary history of organismal lineages is preserved
through the consistent phylogenies derived from a number of core
subsystems including translation, transcription, and DNA
replication~\cite{woese1977phylogenetic,brochier2005emerging,chia2010evolution}.
These subsystems are highly-conserved and present in every cell.
The consistent pattern of relationships within these major cellular
subsystems defines  microbial taxonomy. HGT can then be
detected as gene relationships that differ greatly from this canonical
organismal phylogeny. Horizontal movement of genetic material is
detected by comparing the evolutionary relationships between genes in
different species against their organismal relationships. This can be
done using a number of metrics including correlations in sequence
distance~\cite{farahi2004detection},
phylogeny~\cite{beiko2005highways,kunin2005net}, or gene
composition~\cite{tsirigos2005new,davis2010modal}.

It is noteworthy that many of these methods rely on properly
characterizing the statistical distributions of gene
properties~\cite{farahi2004detection,tsirigos2005new,davis2010modal}.
HGT detection methods generally rely on there being significant
differences in organismal and gene pattern of descent. Limitations on
resolution within the organismal phylogeny, or evolutionary
relationships, makes it difficult to detect HGT between closely related
species. Thus, HGT is nearly undetectable between the organisms for
which it is expected to occur the most frequently. Since we lack the
resolution to distinguish organismal relationships between closely
related species, we cannot track their HGTs through sequence analysis.
Theoretical models of the role of HGT in biology becomes all the more
important for discerning both the extent and possible effects of HGT on
organismal evolution.

What sequence analysis cannot see in nature, fluorescence microscopy
has managed to visualize in the laboratory. By tracking an
\textit{Escherichia coli} plasmid with a fluorescent marker, Babic
\textit{et al.} were able to directly visualize
HGT~\cite{babic2008direct}. While these studies do not tell us much
about the general rate of HGT between species, they can measure the
level of activity of particular MGEs, providing us with a general idea
of what the limits of HGT might be.

HGT may also be detectable indirectly through its influence on genome
organization.  Genomes are not unstructured chains of genes, but
apparently possess an architecture that, particularly in the case of
microbes, can assist in gene expression and genome evolution.  One of
the most frequently-encountered structures in biology is modularity: a
complex network (e.g. metabolic or gene regulatory) that can be
decomposed into independent (i.e. weakly-interacting)
internally-connected functional parts that can evolve separately with
minimal disruption to the system as a
whole\cite{SIMO62,HART99,WAGN07,KREI08}. Modular networks can arise
when the environment is fluctuating in time, creating a
modularly-varying potential for the dynamics\cite{KASH05}, or more
generally, selecting for the organizational structure that can change
in the most facile manner\cite{PART07}.  For a similar reason, a
spatially-heterogeneous environment, such as might arise after an
extinction event, can also promote the emergence of
modularity\cite{kashtan2009extinctions} as a suddenly vacated
ecological niche becomes available for colonization.  One way for
modularity to arise is through the horizontal transfer of a gene or
collection of nearby genes that code for a particular part of a
network\cite{sun07,HE09,DEEM2010}, thereby accelerating the dynamics.
There is evidence that networks can indeed grow by acquiring genes in
groups (known as operons, known to govern coupled reactions in the
cell), and that these are attached preferentially at the edges of the
existing network\cite{PAL05}.

\section{Modeling horizontal gene transfer in biology using statistical mechanics}\label{modeling}

MGEs such as viruses outnumber microbes by over 10-to-1 in
environmental samples~\cite{suttle2005viruses,weinbauer2004ecology}---a
fact suggestive of their larger role in microbial
ecology~\cite{suttle2007marine} and
evolution~\cite{gogarten2005horizontal,frost2005mobile}. Assessing the
role of these MGEs in the ecosystem, however, is difficult. Doing so by
direct experimental measurements, such as sequencing, is essentially
impossible.  Even if one knew every environmental DNA sequence and
biochemical reaction within an environment across time, the task of
reconstructing the effect of HGT on the entire ecosystem would be akin
to calculating the structural soundness of a skyscraper on the basis of
the positions and properties of its elemental particles.

The importance of studying a problem at an appropriate scale applies to
biology as well as physics. The goal of ecology is to understand the
principles underlying change and stability of populations. For example,
the role of a microbial species is something of consequence, whereas
the role of an individual microbe is not. Ecological modeling seeks to
answer questions about the general nature of local processes that give
rise to global behaviors or properties. Understanding how local and
global processes are related can give us an outline of the forces
driving the system.

There are a number of examples in the literature of insight gained from
coarse-graining biological systems, studied using statistical mechanics
ideas ranging from spin glass models\cite{sun07,HE09} to
non-equilibrium statistical mechanics. An example of the latter is the
phenomenon of speciation.  For example, analysis of the codon usage of
genes extracted from libraries of \textit{Escherichia coli} strains
indicate that speciation may have arisen as a result of
HGT~\cite{medigue1991evidence}. A statistical mechanical model shows
specifically how HGT can give rise to speciation---global genome
sequence divergence---in a population of closely-related organisms by
seeding the propagation of mutational
fronts~\cite{vetsigian2005global}.  In this study, statistical
mechanics was used to model the system-wide consequences of the
interplay between point mutation and homologous recombination,
following a single HGT event in microbes.  In asexual organisms, such
as bacteria, homologous recombination allows genome sequences to be
repaired and thus made more uniform in a population, while point
mutations are a source of genome disorder. Species-specific biological
details are important, because the successful insertion of a piece of
alien DNA in a pre-existing bacterial genome relies on the ends of the
insertion matching with both surrounding pre-existing base pairs.  This
requirement of matching ends is absent in some bacteria, or only
enforced at one end in others.  In addition, the cell has mechanisms to
prevent the insertion of alien DNA, but these mechanisms become less
effective the closer the alien DNA sequence matches the region it
replaces.  The behavior of such a population of interacting genomes can
be explored by Monte Carlo simulation, and as a function of the rates
of point mutation and recombination, the phase diagram mapped out.
Interestingly there is a generically first-order phase transition
between a state with a monodisperse population and a state with a
diverse population.  The transition occurs through the propagation of
what have come to be known as diversification fronts propagating along
the genome over the course of evolutionary time.  The front propagation
arises because around an insertion, the disruption of the canonical
sequence means that recombination is locally suppressed, leading to the
build-up of point mutations and the extension of the region of sequence
divergence.  Such fronts are predicted to occur in strains of
\textit{Bacillus cereus} but not in strains of \textit{Buchnera
aphidocola}, owing to the details of their mismatch repair mechanisms,
and these predictions were confirmed by comparative genomics studies on
the fully-sequenced genomes of these organisms. This mechanism for
speciation would leave behind a genome that has a mosaic structure,
corresponding to the merging of several diversification fronts arising
from distinct horizontal gene transfer events.  Indeed, such puzzling
genome features have been observed in an environment where naively one
would have expected extreme selection to have provided essentially no
diversity\cite{TYSO04}.  The mechanism discussed here is especially
interesting, being a counterexample to the popular notion that
speciation arises purely as a result of Darwinian selection.

Other features have been linked to HGT through modeling, including
modularity~\cite{sun07}, the optimality of the genetic
code~\cite{vetsigian2006collective}, and phase variation of
biofilms~\cite{chia2008collective} (a microbial analogue of
multicellular differentiation). Population and gene heterogeneity
within an environment is counterbalanced in microbial systems by
mechanisms such as homologous recombination that serve as an additional
homogenizing force~\cite{wilmes2009dynamic}.

HGT impacts the safety of the biotechnology industry greatly.
Monitoring and modeling the spread of genes such as virulence factors
enhances public safety and helps the development of better lab
practices~\cite{nielsen2004monitoring}. In addition, MGEs can have
impact beyond the movement of genes between organisms. In oceanic
carbon cycling, viruses are thought to ``kill the winner'' since having
the most available hosts should positively feedback into greater
predation~\cite{suttle2007marine}. This dynamic may play an important
role in the diversification of ocean microbes by removing dominant
species~\cite{weinbauer2004viruses} and in the boom-bust cycles of
planktonic blooms, some of which can extend for hundreds of square
miles~\cite{schoemann2005phaeocystis}. Both questions are important to
biodiversity and ecological stability. Plankton blooms choke off oxygen
and nutrients within a large portion of the ocean and may be the result
of trace minerals or other man-made
pollutants~\cite{hallegraeff1993review,beman2005agricultural}. The
factors regulating these boom-bust cycles are a topic of current
debate~\cite{beltrami1994modeling,huppert2002model,menge2009dangerous}.

\section{Horizontal gene transfer and genome entropy}

Quantitative modeling shows that the early benefit from HGT can explain
certain general properties of biology including the emergence of a
universal genetic code~\cite{vetsigian2006collective} and
modularity~\cite{sun07}. Modern life involves a complex
web of enzymatic interactions bringing additional interaction and
regulatory barriers to HGT. However, examples of recent beneficial HGT
are abundant and reveal that HGT is still occurring in modern
organisms. HGT serves as an effective means for modulating mutation
rate within an organism. Modeling shows that mutation rate is indeed
selectable~\cite{denamur2006evolution}. Examples of organisms altering
their mutation rates include the SOS response in \textit{Esherichia
coli}~\cite{mckenzie2000sos} and the competence response of
\textit{Bacillus subtilus}~\cite{hecht2007correlated}.


HGT accelerates organismal evolution by allowing for the exchange of
genes between two organisms, populations, or species. HGT modifies the
types of genetic changes and enhances the amount of total genetic
mutation. In that sense, HGT can be thought of as a source of disorder
in a genome, in effect raising its information entropy~\cite{adami2004information};
thus, we will
informally use the notion of a genome ``temperature" to represent the
level of disorder in a genome. Analogies between HGT and temperature
have been made in the context of studies of the evolution of biological
complexity using digital organisms~\cite{adam00}, and also in the
evolution of cells and genomic annealing~\cite{woese2002evolution}.
Genomic entropy roughly correlates to the amount of genomic change per
unit time. This directly affects both the rate of information loss and
organismal adaptation.  Genomic annealing refers to the concept that
early forms of life had few barriers and much to gain from HGT. This
resulted in massive HGT that slowed down or became quenched as barriers
arose due in part to the increasing complexity of cellular life.

At first glance, these two aspects of genome ``temperature'' appear to
be quite different. However, they can be understood as different sides
of the same coin. Genomic entropy is a property that reflects environmental
information. For success, changes to the environment must be met with
organismal adaptation. In other words, the nature of genome plasticity
must be reflective of fluctuations in the environment. At the same time,
not all aspects of the environment are in constant flux. Ideally a genome
would keep well-adapted genomic elements constant and only change those
that need to change. This is the concept that leads to genomic annealing,
whereby barriers to entropy (in the form of HGT) arise from increasing
complexity. Here we will not attempt
to address issues of how to define complexity; instead, we will attempt
to see if the concept of genome entropy in the information sense can be
helpful phenomenologically.

Competition and changing environments dictate that organisms
that can quickly adapt to these ever changing circumstances will hold
an advantage over their neighbors. It seems natural that a readily
available adaptive mechanism such as HGT would be utilized for exactly
these reasons.

While the frequency of HGT in environmental microbes is inaccessible
through direct measures, the importance of maintaining evolutionary
``temperature'' can be inferred from metagenomic surveys of the
ocean~\cite{konstantinidis2009comparative}. Kostantinidis \textit{et
al} sequenced over 200 Mbp of a random whole-genome shotgun (WGS)
library obtained from a depth of 4,000 m at Station ALOHA in the
Pacific Ocean. The per-bp density of a typical type of transposon known
as an insertion sequence (IS), which can be identified by genes that
codes for Transposase, was measured and compared to that of other
available WGS sequence data from ocean water at various depths.
Overall, they found transposon density increases with ocean depth and
proposed a relaxation of purifying selection at deeper ocean depths
allows the proliferation of these `selfish' gene
elements~\cite{konstantinidis2009comparative}. In other words, the
transposons are simply unchecked by negative (purifying) selection.
Although they are viewed as deleterious, they are allowed to grow in
number due to the lack of competition between organisms. However,
nutrients are scarce in the deep ocean and energy is at a premium: in
such situations, organisms tend toward more efficient genomes rather
than allow them to become disordered. In fact, an explanation based on
the assumed role of purifying selection highlights another interesting
and important problem in marine microbial ecology: what is the source
of genetic diversity in microbial populations?  Naively, one would
expect that in the narrowly-defined environmental niche of the ocean
where there is a limited supply of nutrient and light, the number of
coexisting species, in equilibrium, cannot be greater than the number
of limiting constraints\cite{HUTCH61}.  Observations are in sharp
disagreement with this selection-based idea, and the contradiction has
been a vigorous source of debate in the biological literature.  Amongst
possible proposed solutions to the paradox are spatial and temporal
environmental variability\cite{SCHEF03} and phage
predation\cite{ROHW09,ROHW10}, but a satisfying quantitative
explanation is still lacking.

We propose here an alternative explanation for the apparent increase in
transposons with ocean depth, one tied to the notion of evolutionary
temperature. HGT provides a means of rapid evolution, both increasing
the overall mutation rate and transferring functional genetic material
between organisms. Examples of viruses encoding important functional
genes in the ocean include cyanophage~\cite{sullivan2003cyanophages}
and the photosystem II core reaction~\cite{lindell2004transfer}. As
prevalent as this may seem in the Epipelagic zone of the ocean,
microbial population densities decrease inversely with ocean depth soon
thereafter~\cite{konstantinidis2009comparative}. With reduced average
density, the opportunities for HGT also decrease. On the other hand,
transposons shuffle genetic material within a cell, and this serves to
enhance organismal adaptability.  One might anticipate, therefore, that
to maintain a collection of interacting genomes at the same
temperature, or equivalently, at the same level of disorder,
contributions from all varieties of MGE need to be included.  In
particular, as the HGT rate goes down, the density of other MGEs,
principally transposons would be expected to increase to compensate.

This argument can be translated into a rough scaling argument, as
follows.  The number density $\rho$ of cells is a decreasing function
of depth, falling off roughly as $\rho \sim d^{-1}$ where $d$ is the
depth in the ocean, according to the
data\cite{konstantinidis2009comparative}. The probability of HGT events
per base pair, either conjugal or mediated by an intermediary, $P_{HGT}
\sim \rho^2$, assuming some sort of law of mass action.  On the other
hand, transposons shuffle genes within a cell's genome, and this
process actually requires two IS elements in order for the tranposition
event to take place along the genome, so that the probability of each
tranposition event is proportional to the square of the number density
of IS elements, $\rho_{IS}$.  Making the assumption that there is a
uniform genome entropy with depth, and that the HGT and transposition
events are independent, we obtain that
\begin{equation}
P_{HGT} \times \rho_{IS}^2 \sim \hbox{constant} < 1.
\end{equation}
Making the usual sort of mean field approximation, we
then obtain that
\begin{equation}
\rho^2 \times \rho_{IS}^2 \sim \hbox{constant}
\end{equation}
and thus that $\rho_{IS} \sim d$, a result in rough agreement with the
available data.

\begin{figure}
  \includegraphics[width=\columnwidth]{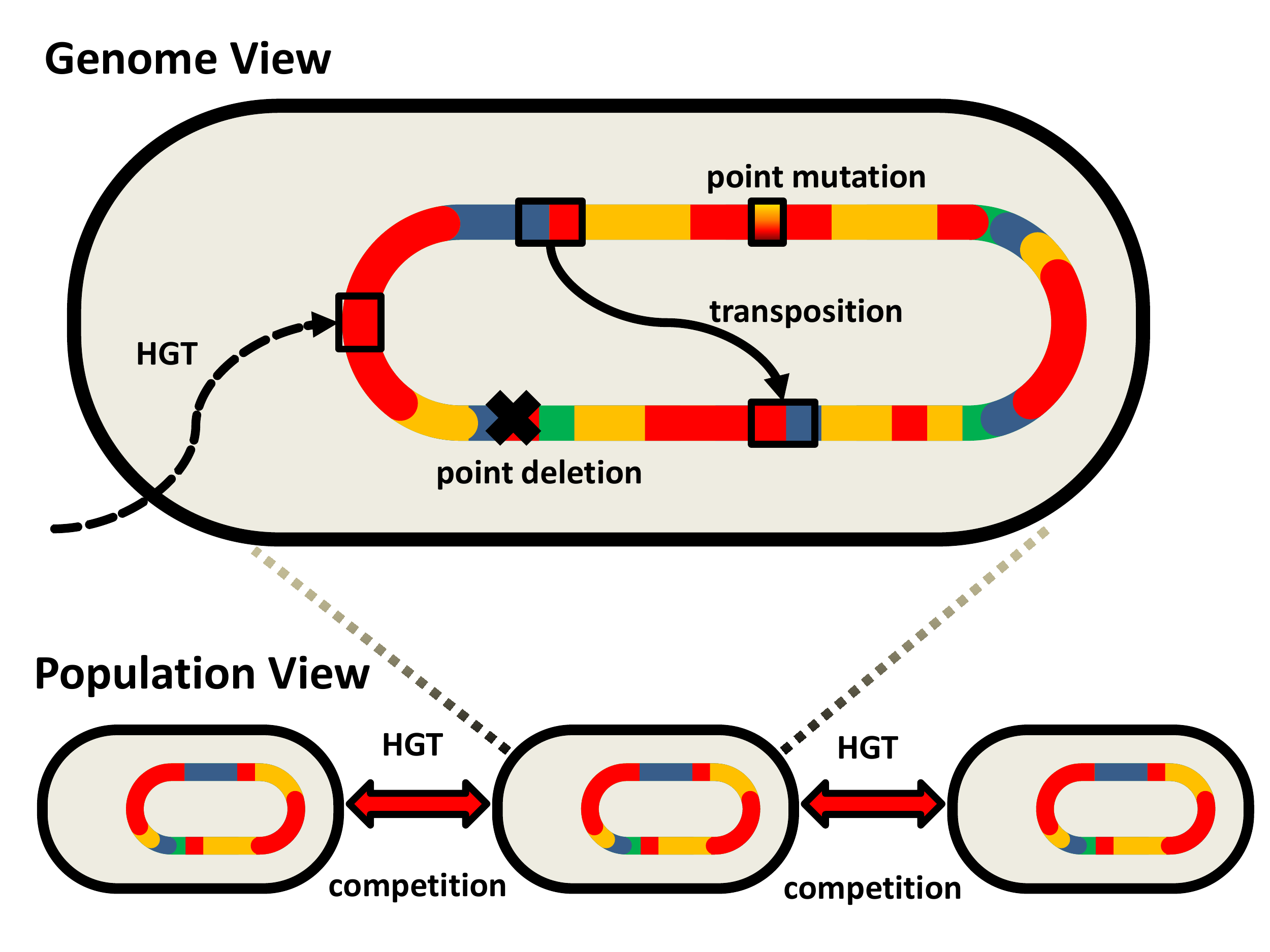}
\caption{Schematic of processes present in simulation}
\label{fig:HGT}       
\end{figure}

\begin{figure}
  \includegraphics[width=\columnwidth]{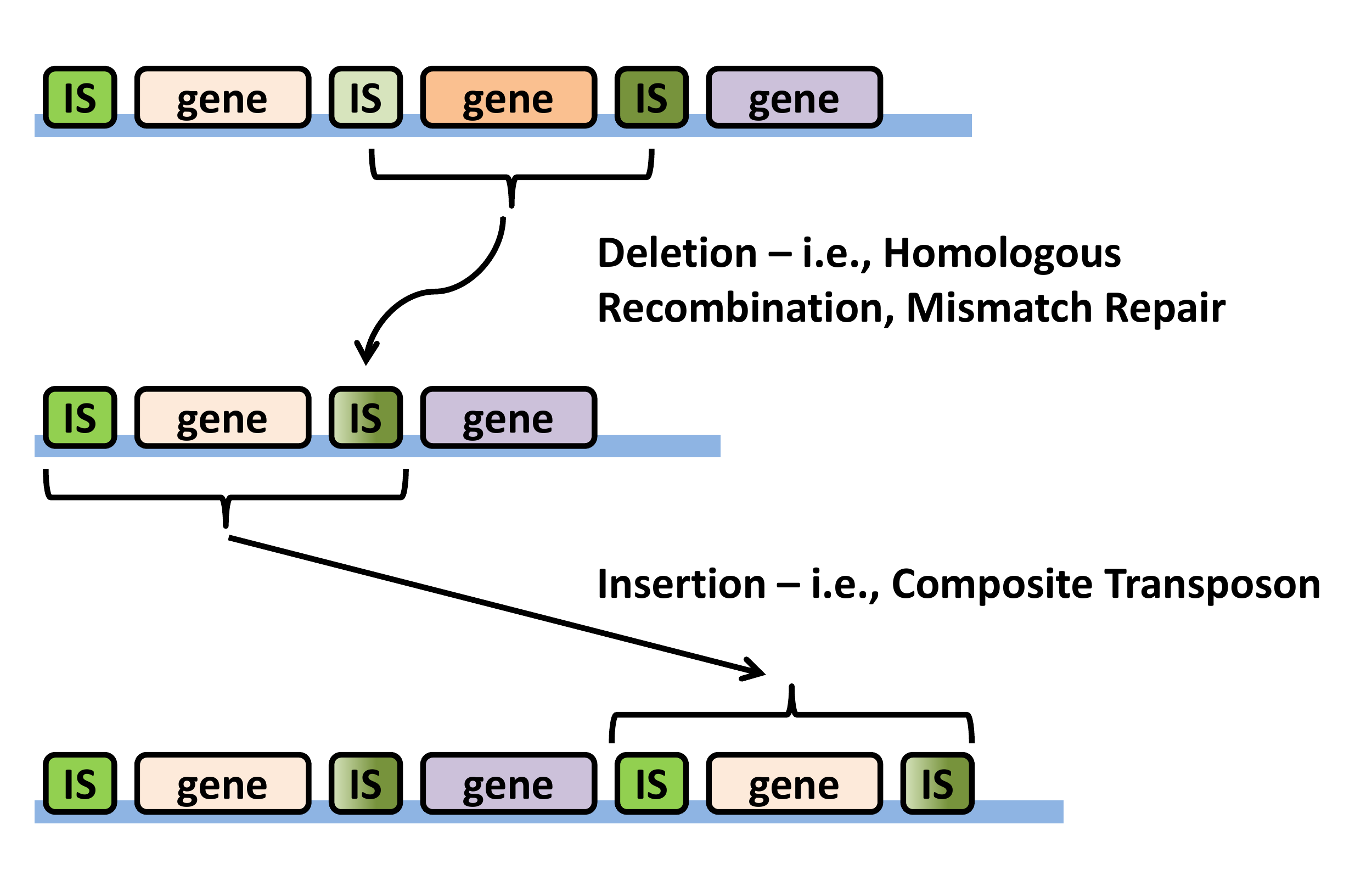}
\caption{Schematic of how insertion sequences (IS) act to transpose and delete
genetic material.}
\label{fig:IS}       
\end{figure}

In order to test this idea, we simulated populations of microbes under
continuous selection pressure for a fixed target sequence at different
population densities. A schematic of the the processes that impact our
simulations is given by Fig.~\ref{fig:HGT}.
For initial conditions, we take genomes of length
300 whose letters are selected randomly from an alphabet size of 20.
These genomes are each initialized randomly and a variable number $N$
of them are placed into an array of size 10000. The population in each
subset then remains fixed, but competition is allowed whereby a fitter
organism may replace a less fit organism. Point mutation, deletion,
HGT, and transposons are allowed. Point mutation randomly exchanges one
letter of the genome for another randomly assigned letter while
deletion removes the letter entirely. HGT randomly selects a segment 10
letters long from one genome and inserts it into another. The model for
transposon behavior is based on Insertion Sequence (IS) element
behavior. In this model, an IS is represented by a specific 2 letter
combination. As shown in Fig~\ref{fig:IS}, an IS can non-conservatively
insert (i.e., copy-paste as
opposed to cut-paste) itself elsewhere within a genome. When 2 IS
elements are within a fixed distance of each other (20 in this
simulation), then they either transpose the entire length of genome in
between including themselves or the region between them is deleted
through homologous recombination, each event occurring with equal
probability. The rates of mutation, deletion, and transposition are
then fixed at 0.001, 3.0, and 0.5 per generation per organism. The rate
of HGT is fixed at 1.0 per opportunity. In order for HGT or competition
(whereby one organism overwrites another) to occur, two slots in the
population array are selected randomly. If both slots contain a living
organism, then HGT occurs according to the HGT rate. In this way, as
population decrease, so does the number of HGT opportunities. The
parameters were chosen with an eye toward obtaining the correct relative
rates for each of the processes, as mimicking the orders of magnitude
difference between mutation rate and HGT ($10^9$) becomes impractical
computationally. We examined mutation rate and system size scaling in
for a related system and found that our results were qualitatively
robust~\cite{chia2010dynamics}.

Genes within the genome are identified as a gapless subset of the
genome that shares a common subsequence with the target sequence, which
is of length 10. The organismal fitness is then given by
\begin{equation}
F = \min(\sum_{i=0}^{n} g_i,1) - n \mu
\label{eq:F}
\end{equation}
where $g_i$ is the length of the $i$th gene (note that, here, length relates directly to the
number of consecutive letters that match the target sequence), $n$ is the total
number of genes within a genome, and $\mu$ is a penalty of 0.2 for each gene copy
present in the genome. This fitness rewards genomes containing
subsequences matching a particular target sequence. To some extent,
there can be a fitness gain from multiple copies of a gene, but
this gain is not limitless and many short matches to the target sequence
are penalized. The fitness function described here has a basis in a
recently proposed model for the evolution of novel gene function~\cite{bergthorsson2007osd}.
A similar fitness function was also at the basis
of related theoretical work on the transposon dynamics of recent
obligate associations~\cite{chia2010dynamics}.

\begin{figure}
  \includegraphics[width=\columnwidth]{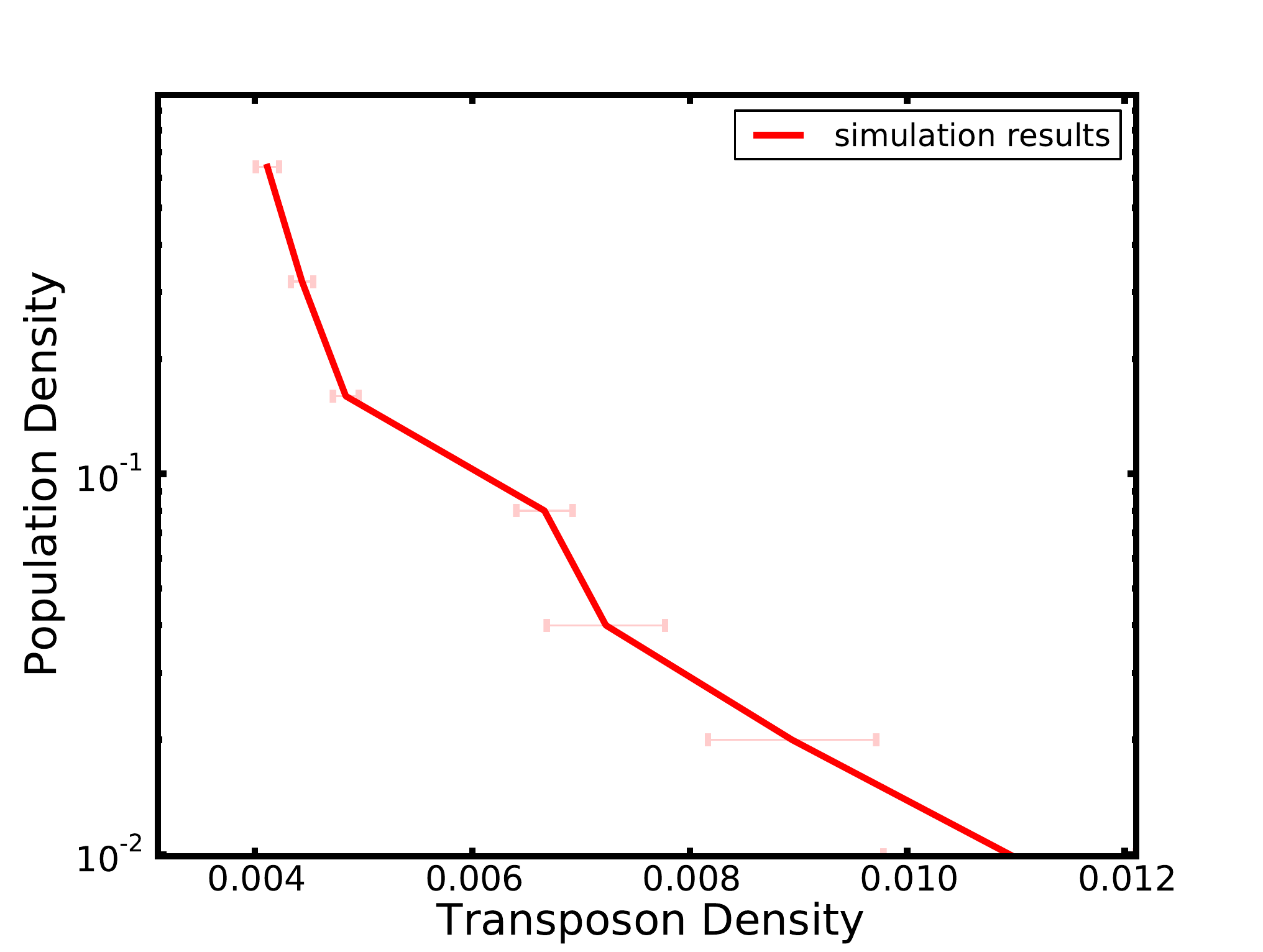}
\caption{Transposon density versus population density. We simulated the
adaptation of organisms toward a fixed target sequence allowing point
mutation, deletion, HGT, and transposon dynamics for 20,000 generations
and plot the averaged final transposon densities of 100 simulations for
each point. Different population densities have differing amounts of
opportunities for HGT. Our simulation results show that at lower
population densities with decreased HGT transposons increase in
relative abundance. This is consistent with the idea that transposons
serve as a substitute for the evolutionary dynamic provided by HGT in
shallower waters. In ocean waters, population density is inversely
proportional to depth below the Epipelagic Zone ($>$200 m), to first
order approximation~\cite{konstantinidis2009comparative}.}
\label{fig:1}       
\end{figure}

Fig.~\ref{fig:1} plots the results from our simulation of microbial
competition according to the rules outlined above across a population
density gradient. This tests the hypothesis that the apparent increase
in transposon density is related to decreasing levels of HGT due to the
decreasing population densities in deeper waters. As shown, we do
indeed see an increase in the transposon density that corresponds with
decreasing population density. However, in order to assert that this is
due to the lack of HGT and not other effects such as population size or
competition timescales (which have also slowed down due to competition
requiring two organisms to interact), we must isolate these other
effects and focus on HGT alone.

\begin{figure}
  \includegraphics[width=\columnwidth]{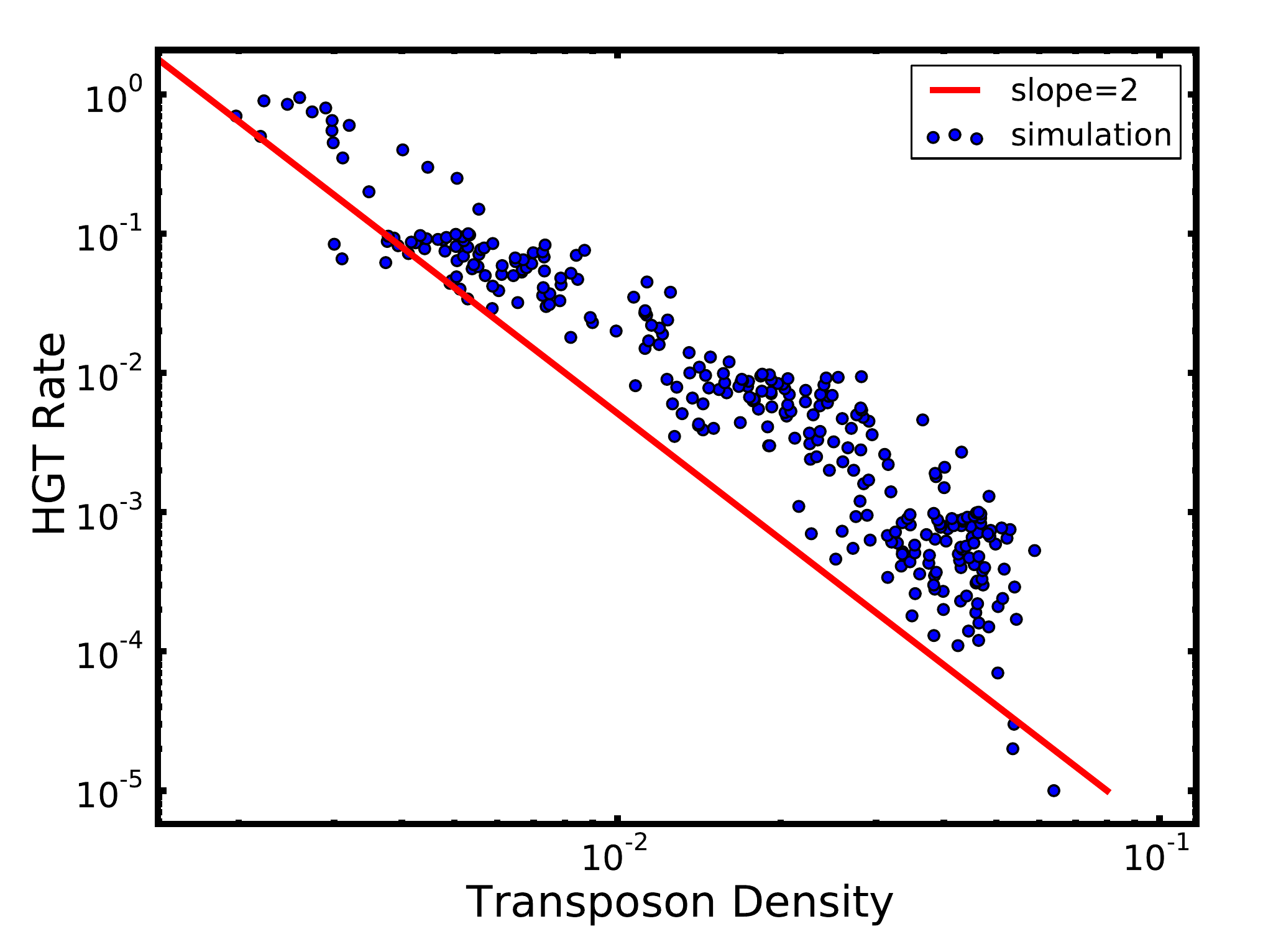}
\caption{Transposon density versus HGT rate for fixed population size.
Keeping population size at the fixed value of 1,000, we plot the
transposon density after 10,000 generations averaged across 100
simulations per point. The deletion rate is 2 per generation per
organism. All other parameters are the same as described in the main
text. In order to verify that the increase in transposon density seen
in Fig.~\ref{fig:1} is the result of an HGT:transposon evolutionary
tradeoff, we eliminate other sources of variation by fixing the
population size and varying the HGT rate.  This serves to eliminate the
effects of differing population sizes and competition timescales
present in the previous simulation. In this simulation, we show
increases in transposon density with decreasing HGT rate. We plot a
power law line with an exponent of 2 for reference. The line does not
represent a fit to the data.}
\label{fig:2}       
\end{figure}

Fig.~\ref{fig:2} shows what happens when we allow a fixed population
size to adapt under different HGT rates. The trend of increasing
transposon density holds for decreasing HGT rate. This is consistent
with our hypothesis that the transposon density trend seen in the ocean
corresponds to lower HGT rates in the ocean depths. Reduction of HGT
leads to increased transposon numbers as originally postulated.

The possibility that some notion of `adaptability' underlies transposon
dynamics points toward two very interesting ideas. The first is that
transposons, as well as other mobile genetic elements such as viruses
and plasmids, could be more than just parasitic gene sequences feeding
off host genomes. Instead, they may be required for the long term
survival, adaptability, and diversification of an organismal lineage.
The second is that evolutionary dynamics such as transposon
proliferation may be driven by generic processes rather than governed
by the specific histories of individual populations. This means that
the properties of biological organism can and should be understood from
the viewpoint of the statistics of a physical process rather than the
particulars of a historical accident.

\section{Summary}

HGT couples together unrelated organismal lineages in a way that
requires us to rethink the classical point-mutation based notions of
population genetics. HGT has been shown to be a prevalent force in
microbial evolution especially. The impact of HGT on gene and
organismal evolution has not been fully understood. However, notions
from statistical physics such as temperature and annealing have been
applied to evolutionary dynamics before and will no doubt continue to
play a role in determining the underlying rules of evolution. We showed
one example of how theory can play a role by offering up a quantitative
hypothesis on the role of HGT in determining how transposon densities
vary with ocean depth. Our example involves very generic interactions
and does not depend on the particular environment or microbial species.
It is exactly these non-specific types of studies that are important if
we are to unify our understanding of the dynamics of genome evolution.

\begin{acknowledgements}
We thank Ed DeLong, Carl Woese and Nicholas Guttenberg for valuable
discussions, and Pan-Jun Kim and Zhenyu Wang for helpful comments on
the manuscript. This work was partially supported by the National
Science Foundation through grant NSF-EF-0526747.
\end{acknowledgements}

\bibliographystyle{spmpsci}      
\bibliography{stathgt}   

\begin{thebibliography}{10}
\providecommand{\url}[1]{{#1}}
\providecommand{\urlprefix}{URL }
\expandafter\ifx\csname urlstyle\endcsname\relax
  \providecommand{\doi}[1]{DOI~\discretionary{}{}{}#1}\else
  \providecommand{\doi}{DOI~\discretionary{}{}{}\begingroup
  \urlstyle{rm}\Url}\fi

\bibitem{adami2004information}
Adami, C.: {Information theory in molecular biology}.
\newblock Physics of Life Reviews \textbf{1}(1), 3--22 (2004)

\bibitem{adam00}
Adami, C., Ofria, C., Collier, T.: {Evolution of biological complexity}.
\newblock Proc. Natl. Acad. Sci. USA \textbf{97}(9), 4463--4468 (2000)

\bibitem{aminov2007evolution}
Aminov, R.I., Mackie, R.I.: Evolution and ecology of antibiotic resistance
  genes.
\newblock FEMS Microbiol. Lett. \textbf{271}(2), 147--161 (2007)

\bibitem{ANDE66}
Anderson, E.S.: Possible importance of transfer factors in bacterial evolution.
\newblock Nature \textbf{209}, 637--638 (1966)

\bibitem{ANDE70}
Anderson, N.G.: Evolutionary significance of virus infection.
\newblock Nature \textbf{227}, 1346--1347 (1970)

\bibitem{babic2008direct}
Babic, A., Lindner, A.B., Vulic, M., Stewart, E.J., Radman, M.: Direct
  visualization of horizontal gene transfer.
\newblock Science \textbf{319}(5869), 1533--1536 (2008)

\bibitem{beiko2005highways}
Beiko, R.G., Harlow, T.J., Ragan, M.A.: Highways of gene sharing in
  prokaryotes.
\newblock Proc. Natl. Acad. Sci. U.S.A. \textbf{102}(40), 14,332--14,338 (2005)

\bibitem{beltrami1994modeling}
Beltrami, E., Carroll, T.: Modeling the role of viral disease in recurrent
  phytoplankton blooms.
\newblock J. Math. Biol. \textbf{32}(8), 857--863 (1994)

\bibitem{ROHW10}
Beltran, R., LinLin~Li, L., Mike~Furlan, F., Mya~Breitbart, J.,
  Christelle~Desnues, E., Robert~Edwards, B., Matthew~Haynes, H., David~Lipson,
  J., Anna Belen Martin-Cuadrado, A., Jim~Nulton, L., et~al.: {Viral and
  microbial community dynamics in four aquatic environments}.
\newblock The ISME Journal \textbf{4}(6), 739--751 (2010)

\bibitem{beman2005agricultural}
Beman, J.M., Arrigo, K.R., Matson, P.A.: Agricultural runoff fuels large
  phytoplankton blooms in vulnerable areas of the ocean.
\newblock Nature \textbf{434}(7030), 211--214 (2005)

\bibitem{bennett2008plasmid}
Bennett, P.M.: Plasmid encoded antibiotic resistance: acquisition and transfer
  of antibiotic resistance genes in bacteria.
\newblock Brit. J. Pharm. \textbf{153}(S1), S347--S357 (2008)

\bibitem{bergthorsson2007osd}
Bergthorsson, U., Andersson, D.I., Roth, J.R.: Ohno's dilemma: Evolution of new
  genes under continuous selection.
\newblock Proc. Natl. Acad. Sci. U.S.A. \textbf{104}(43), 17,004 (2007)

\bibitem{brochier2005emerging}
Brochier, C., Forterre, P., Gribaldo, S.: An emerging phylogenetic core of
  archaea: phylogenies of transcription and translation machineries converge
  following addition of new genome sequences.
\newblock BMC Evol. Biol. \textbf{5}(1), 36 (2005)

\bibitem{butler2009robust}
Butler, T., Goldenfeld, N.: Robust ecological pattern formation induced by
  demographic noise.
\newblock Phys. Rev. E \textbf{80}(3), 30,902 (2009)

\bibitem{chen2004dna}
Chen, I., Dubnau, D.: {DNA} uptake during bacterial transformation.
\newblock Nat. Rev. Microbiol. \textbf{2}(3), 241--249 (2004)

\bibitem{chia2010evolution}
Chia, N., Cann, I., Olsen, G.J.: {Evolution of DNA Replication Protein
  Complexes in Eukaryotes and Archaea}.
\newblock PLoS ONE \textbf{5}(6), e10,866 (2010)

\bibitem{chia2010dynamics}
Chia, N., Goldenfeld, N.: The dynamics of gene duplication and transposons in
  microbial genomes following a sudden environmental change.
\newblock Arxiv preprint arXiv:1005. 3349  (2010)

\bibitem{chia2008collective}
Chia, N., Woese, C.R., Goldenfeld, N.: A collective mechanism for phase
  variation in biofilms.
\newblock Proc. Natl. Acad. Sci. U.S.A. \textbf{105}(38), 14,597--14,603 (2008)

\bibitem{cohen2010inference}
Cohen, O., Pupko, T.: Inference and characterization of horizontally
  transferred gene families using stochastic mapping.
\newblock Mol. Biol. Evol. \textbf{27}(3), 703 (2010)

\bibitem{davis2010modal}
Davis, J.J., Olsen, G.J.: Modal codon usage: Assessing the typical codon usage
  of a genome.
\newblock Mol. Biol. Evol. \textbf{27}(4), 800 (2010)

\bibitem{de2003proteorhodopsin}
De~La~Torre, J.R., Christianson, L.M., B{\'e}j{\`a}, O., Suzuki, M.T., Karl,
  D.M., Heidelberg, J., DeLong, E.F.: Proteorhodopsin genes are distributed
  among divergent marine bacterial taxa.
\newblock Proc. Natl. Acad. Sci. U.S.A. \textbf{100}(22), 12,830 (2003)

\bibitem{delong2005microbial}
DeLong, E.: {Microbial community genomics in the ocean}.
\newblock Nature Reviews Microbiology \textbf{3}(6), 459--469 (2005)

\bibitem{denamur2006evolution}
Denamur, E., Matic, I.: Evolution of mutation rates in bacteria.
\newblock Mol. Microbiol. \textbf{60}(4), 820--827 (2006)

\bibitem{dewar2008statistical}
Dewar, R.C., Porte, A.: Statistical mechanics unifies different ecological
  patterns.
\newblock J. Theor. Biol. \textbf{251}(3), 389--403 (2008)

\bibitem{ROHW05}
Edwards, R., Rohwer, F.: {Viral metagenomics}.
\newblock Nature Reviews Microbiology \textbf{3}(6), 504--510 (2005)

\bibitem{eisen2000horizontal}
Eisen, J.A.: Horizontal gene transfer among microbial genomes: new insights
  from complete genome analysis.
\newblock Curr. Opin. Genet. Dev. \textbf{10}(6), 606--611 (2000)

\bibitem{elsas2002ecology}
Elsas, J.D., Bailey, M.J.: The ecology of transfer of mobile genetic elements.
\newblock FEMS Microbiol. Ecol. \textbf{42}(2), 187--197 (2002)

\bibitem{farahi2004detection}
Farahi, K., Pusch, G.D., Overbeek, R., Whitman, W.B.: Detection of lateral gene
  transfer events in the prokaryotic trna synthetases by the ratios of
  evolutionary distances method.
\newblock J. Mol. Evol. \textbf{58}(5), 615--631 (2004)

\bibitem{frigaard2006proteorhodopsin}
Frigaard, N.U., Martinez, A., Mincer, T.J., DeLong, E.F.: Proteorhodopsin
  lateral gene transfer between marine planktonic {B}acteria and {A}rchaea.
\newblock Nature \textbf{439}(7078), 847--850 (2006)

\bibitem{frost2005mobile}
Frost, L.S., Leplae, R., Summers, A.O., Toussaint, A.: Mobile genetic elements:
  the agents of open source evolution.
\newblock Nat. Rev. Microbiol. \textbf{3}(9), 722--732 (2005)

\bibitem{garcia2000horizontal}
Garcia-Vallv{\'e}, S., Romeu, A., Palau, J.: Horizontal gene transfer in
  bacterial and archaeal complete genomes.
\newblock Genome Res. \textbf{10}(11), 1719 (2000)

\bibitem{gladyshev2008massive}
Gladyshev, E., Meselson, M., Arkhipova, I.: {Massive horizontal gene transfer
  in bdelloid rotifers}.
\newblock Science \textbf{320}(5880), 1210 (2008)

\bibitem{gogarten2005horizontal}
Gogarten, J.P., Townsend, J.P.: Horizontal gene transfer, genome innovation and
  evolution.
\newblock Nat. Rev. Microbiol. \textbf{3}(9), 679--687 (2005)

\bibitem{goldenfeld2007biology}
Goldenfeld, N., Woese, C.: Biology's next revolution.
\newblock Nature \textbf{445}(7126), 369 (2007)

\bibitem{GOLD10}
Goldenfeld, N., Woese, C.: Life is physics: evolution as a collective
  phenomenon far from equilibrium.
\newblock Ann. Rev. Cond. Matt. Phys. \textbf{1}, in press (2010)

\bibitem{hallegraeff1993review}
Hallegraeff, G.: A review of harmful algal blooms and their apparent global
  increase.
\newblock Phycologia \textbf{32}(2), 79--99 (1993)

\bibitem{handelsman2004metagenomics}
Handelsman, J.: {Metagenomics: application of genomics to uncultured
  microorganisms}.
\newblock Microbiology and Molecular Biology Reviews \textbf{68}(4), 669--685
  (2004)

\bibitem{HART99}
Hartwell, L., Hopfield, J., Leibler, S., Murray, A.: {From molecular to modular
  cell biology}.
\newblock Nature \textbf{402}, C47--C52 (1999)

\bibitem{HE09}
He, J., Sun, J., Deem, M.: {Spontaneous emergence of modularity in a model of
  evolving individuals and in real networks}.
\newblock Physical Review E \textbf{79}(3), 31,907 (2009)

\bibitem{hecht2007correlated}
Hecht, I., Ben-Jacob, E., Levine, H.: Correlated phenotypic transitions to
  competence in bacterial colonies.
\newblock Phys. Rev. E \textbf{76}(4), 40,901 (2007)

\bibitem{hendrix2000origins}
Hendrix, R.W., Lawrence, J.G., Hatfull, G.F., Casjens, S.: {The origins and
  ongoing evolution of viruses}.
\newblock Trends in Microbiology \textbf{8}(11), 504--508 (2000)

\bibitem{holden2004complete}
Holden, M.T.G., Feil, E.J., Lindsay, J.A., Peacock, S.J., Day, N.P.J., Enright,
  M.C., Foster, T.J., Moore, C.E., Hurst, L., Atkin, R., et~al.: Complete
  genomes of two clinical staphylococcus aureus strains: evidence for the rapid
  evolution of virulence and drug resistance.
\newblock Proc. Natl. Acad. Sci. U.S.A. \textbf{101}(26), 9786--9792 (2004)

\bibitem{hotopp2007wlg}
Hotopp, J., Clark, M., Oliveira, D., Foster, J., Fischer, P., Torres, M.,
  Giebel, J., Kumar, N., Ishmael, N., Wang, S., et~al.: {Widespread lateral
  gene transfer from intracellular bacteria to multicellular eukaryotes}.
\newblock Science \textbf{317}(5845), 1753--1756 (2007)

\bibitem{hubbell2001unified_b}
Hubbell, S.P.: The unified neutral theory of species abundance and diversity.
\newblock Princeton, Princeton University (2001)

\bibitem{huppert2002model}
Huppert, A., Blasius, B., Stone, L.: A model of phytoplankton blooms.
\newblock Amer. Naturalist \textbf{159}(2), 156--171 (2002)

\bibitem{HUTCH61}
Hutchinson, G.: {The paradox of the plankton}.
\newblock American Naturalist \textbf{95}(882), 137--145 (1961)

\bibitem{jeong2000large}
Jeong, H., Tombor, B., Albert, R., Oltvai, Z.N., Barab{\'a}si, A.L.: The
  large-scale organization of metabolic networks.
\newblock Nature \textbf{407}(6804), 651--654 (2000)

\bibitem{KASH05}
Kashtan, N., Alon, U.: {Spontaneous evolution of modularity and network
  motifs}.
\newblock Proceedings of the National Academy of Sciences \textbf{102}(39),
  13,773--13,778 (2005)

\bibitem{kashtan2009extinctions}
Kashtan, N., Parter, M., Dekel, E., Mayo, A., Alon, U.: {Extinctions in
  heterogeneous environments and the evolution of modularity}.
\newblock Evolution \textbf{63}(8), 1964--1975 (2009)

\bibitem{keeling2008horizontal}
Keeling, P.J., Palmer, J.D.: {Horizontal gene transfer in eukaryotic
  evolution}.
\newblock Nat. Reviews Genetics \textbf{9}(8), 605--618 (2008)

\bibitem{koike2007monitoring}
Koike, S., Krapac, I., Oliver, H., Yannarell, A., Chee-Sanford, J., Aminov, R.,
  Mackie, R.: Monitoring and source tracking of tetracycline resistance genes
  in lagoons and groundwater adjacent to swine production facilities over a
  3-year period.
\newblock Appl. Environ. Microbiol. \textbf{73}(15), 4813--4823 (2007)

\bibitem{konstantinidis2009comparative}
Konstantinidis, K.T., Braff, J., Karl, D.M., DeLong, E.F.: Comparative
  metagenomic analysis of a microbial community residing at a depth of 4,000
  meters at station {ALOHA} in the {N}orth {P}acific subtropical gyre.
\newblock Appl. Environ. Microbiol. \textbf{75}(16), 5345--5355 (2009)

\bibitem{KREI08}
Kreimer, A., Borenstein, E., Gophna, U., Ruppin, E.: {The evolution of
  modularity in bacterial metabolic networks}.
\newblock Proceedings of the National Academy of Sciences \textbf{105}(19),
  6976--6981 (2008)

\bibitem{kunin2005net}
Kunin, V., Goldovsky, L., Darzentas, N., Ouzounis, C.A.: The net of life:
  reconstructing the microbial phylogenetic network.
\newblock Genome Res. \textbf{15}(7), 954--959 (2005)

\bibitem{lindell2004transfer}
Lindell, D., Sullivan, M.B., Johnson, Z.I., Tolonen, A.C., Rohwer, F.,
  Chisholm, S.W.: Transfer of photosynthesis genes to and from
  {P}rochlorococcus viruses.
\newblock Proc. Natl. Acad. Sci. U.S.A. \textbf{101}(30), 11,013--110,139
  (2004)

\bibitem{DEEM2010}
Lorenz, D.M., Jeng, A., Deem, M.W.: The emergence of modularity in biological
  systems (2010).
\newblock Submitted to Journal of Life Reviews

\bibitem{lorenz1994bacterial}
Lorenz, M., Wackernagel, W.: Bacterial gene transfer by natural genetic
  transformation in the environment.
\newblock Microbiol. Mol. Biol. Rev. \textbf{58}(3), 563 (1994)

\bibitem{mahillon1998insertion}
Mahillon, J., Chandler, M.: {Insertion sequences}.
\newblock Microbiol. Mol. Biol. Rev. \textbf{62}(3), 725 (1998)

\bibitem{mckenzie2000sos}
McKenzie, G.J., Harris, R.S., Lee, P.L., Rosenberg, S.M.: The sos response
  regulates adaptive mutation.
\newblock Proc. Natl. Acad. Sci. U.S.A. \textbf{97}(12), 6646 (2000)

\bibitem{medigue1991evidence}
M{\'e}digue, C., Rouxel, T., Vigier, P., H{\'e}naut, A., Danchin, A.: Evidence
  for horizontal gene transfer in {E}scherichia coli speciation.
\newblock J. Mol. Biol. \textbf{222}(4), 851--856 (1991)

\bibitem{menge2009dangerous}
Menge, D.N.L., Weitz, J.S.: Dangerous nutrients: Evolution of phytoplankton
  resource uptake subject to virus attack.
\newblock J. Theor. Biol. \textbf{257}(1), 104--115 (2009)

\bibitem{mirollo1990synchronization}
Mirollo, R.E., Strogatz, S.H.: Synchronization of pulse-coupled biological
  oscillators.
\newblock SIAM J. Appl. Math. \textbf{50}(6), 1645--1662 (1990)

\bibitem{monier2009horizontal}
Monier, A., Pagarete, A., de~Vargas, C., Allen, M., Read, B., Claverie, J.,
  Ogata, H.: {Horizontal gene transfer of an entire metabolic pathway between a
  eukaryotic alga and its DNA virus}.
\newblock Genome Research \textbf{19}(8), 1441 (2009)

\bibitem{nielsen2004monitoring}
Nielsen, K.M., Townsend, J.P.: Monitoring and modeling horizontal gene
  transfer.
\newblock Nat. Biotechnol. \textbf{22}(9), 1110--1114 (2004)

\bibitem{pace2008repeated}
Pace, J., Gilbert, C., Clark, M., Feschotte, C.: {Repeated horizontal transfer
  of a DNA transposon in mammals and other tetrapods}.
\newblock Proceedings of the National Academy of Sciences \textbf{105}(44),
  17,023 (2008)

\bibitem{PAL05}
P{\'a}l, C., Papp, B., Lercher, M.: {Adaptive evolution of bacterial metabolic
  networks by horizontal gene transfer}.
\newblock Nature genetics \textbf{37}(12), 1372--1375 (2005)

\bibitem{PART07}
Parter, M., Kashtan, N., Alon, U.: {Environmental variability and modularity of
  bacterial metabolic networks}.
\newblock BMC Evolutionary Biology \textbf{7}, 169/1--8 (2007)

\bibitem{prangishvili2006viruses}
Prangishvili, D., Forterre, P., Garrett, R.A.: Viruses of the archaea: a
  unifying view.
\newblock Nat. Rev. Microbiol. \textbf{4}(11), 837--848 (2006)

\bibitem{ROHW09}
Rodriguez-Valera, F., Martin-Cuadrado, A., Rodriguez-Brito, B., Pasic, L.,
  Thingstad, T., Rohwer, F., Mira, A.: {Explaining microbial population
  genomics through phage predation}.
\newblock Nature Reviews Microbiology \textbf{7}(11), 828--836 (2009)

\bibitem{salyers1997antibiotic}
Salyers, A.A., Amabile-Cuevas, C.F.: {Why are antibiotic resistance genes so
  resistant to elimination?}
\newblock Antimicrobial Agents and Chemotherapy \textbf{41}(11), 2321 (1997)

\bibitem{SCHEF03}
Scheffer, M., Rinaldi, S., Huisman, J., Weissing, F.: {Why plankton communities
  have no equilibrium: solutions to the paradox}.
\newblock Hydrobiologia \textbf{491}(1), 9--18 (2003)

\bibitem{schoemann2005phaeocystis}
Schoemann, V., Becquevort, S., Stefels, J., Rousseau, V., Lancelot, C.:
  Phaeocystis blooms in the global ocean and their controlling mechanisms: a
  review.
\newblock J. Sea Res. \textbf{53}(1-2), 43--66 (2005)

\bibitem{SIMO62}
Simon, H.: {The architecture of complexity}.
\newblock Proceedings of the American Philosophical Society \textbf{106}(6),
  467--482 (1962)

\bibitem{snel2002genomes}
Snel, B., Bork, P., Huynen, M.A.: Genomes in flux: the evolution of archaeal
  and proteobacterial gene content.
\newblock Genome Res. \textbf{12}(1), 17--25 (2002)

\bibitem{sonea1988bwl}
Sonea, S.: {A bacterial way of life}.
\newblock Nature \textbf{331}(6153), 216 (1988)

\bibitem{sorensen2005studying}
S{\o}rensen, S.J., Bailey, M., Hansen, L.H., Kroer, N., Wuertz, S.: {Studying
  plasmid horizontal transfer in situ: a critical review}.
\newblock Nat. Rev. Microbiol. \textbf{3}(9), 700--710 (2005)

\bibitem{sullivan2003cyanophages}
Sullivan, M.B., Waterbury, J.B., Chisholm, S.W.: {C}yanophages infecting the
  oceanic cyanobacterium {P}rochlorococcus.
\newblock Nature \textbf{424}(6952), 1047--1051 (2003)

\bibitem{sun07}
Sun, J., Deem, M.W.: Spontaneous emergence of modularity in a model of evolving
  individuals.
\newblock Physical Review Letters \textbf{99}(22), 228,107/1--4 (2007)

\bibitem{suttle2005viruses}
Suttle, C.A.: Viruses in the sea.
\newblock Nature \textbf{437}(7057), 356--361 (2005)

\bibitem{suttle2007marine}
Suttle, C.A.: Marine viruses�major players in the global ecosystem.
\newblock Nat. Rev. Microbiol. \textbf{5}(10), 801--812 (2007)

\bibitem{syvanen1994hgt}
Syvanen, M.: {Horizontal gene transfer: evidence and possible consequences}.
\newblock Annual Review of Genetics \textbf{28}(1), 237--261 (1994)

\bibitem{thomas2005mechanisms}
Thomas, C.M., Nielsen, K.M.: Mechanisms of, and barriers to, horizontal gene
  transfer between bacteria.
\newblock Nat. Rev. Microbiol. \textbf{3}(9), 711--721 (2005)

\bibitem{thomas1995dynamical}
Thomas, R., Thieffry, D., Kaufman, M.: Dynamical behaviour of biological
  regulatory networks�i. biological role of feedback loops and practical use
  of the concept of the loop-characteristic state.
\newblock Bull. Math. Biol. \textbf{57}(2), 247--276 (1995)

\bibitem{tsirigos2005new}
Tsirigos, A., Rigoutsos, I.: A new computational method for the detection of
  horizontal gene transfer events.
\newblock Nucleic Acids Res. \textbf{33}(3), 922 (2005)

\bibitem{TYSO04}
Tyson, G., Chapman, J., Hugenholtz, P., Allen, E., Ram, R., Richardson, P.,
  Solovyev, V., Rubin, E., Rokhsar, D., Banfield, J.: {Community structure and
  metabolism through reconstruction of microbial genomes from the environment}.
\newblock Nature \textbf{428}(6978), 37--43 (2004)

\bibitem{vetsigian2005global}
Vetsigian, K., Goldenfeld, N.: Global divergence of microbial genome sequences
  mediated by propagating fronts.
\newblock Proc. Natl. Acad. Sci. U.S.A. \textbf{102}(20), 7332 (2005)

\bibitem{vetsigian2006collective}
Vetsigian, K., Woese, C., Goldenfeld, N.: Collective evolution and the genetic
  code.
\newblock Proc. Natl. Acad. Sci. U.S.A. \textbf{103}(28), 10,696--10,702 (2006)

\bibitem{WAGN07}
Wagner, G., Pavlicev, M., Cheverud, J.: {The road to modularity}.
\newblock Nature Reviews Genetics \textbf{8}(12), 921--931 (2007)

\bibitem{BASS05}
Waters, C., Bassler, B.: {Quorum Sensing: Cell-to-Cell Communication in
  Bacteria}.
\newblock Annu. Rev. Cell Dev. Biol \textbf{21}, 319--46 (2005)

\bibitem{weinbauer2004ecology}
Weinbauer, M.G.: Ecology of prokaryotic viruses.
\newblock FEMS Microbiol. Rev. \textbf{28}(2), 127--181 (2004)

\bibitem{weinbauer2004viruses}
Weinbauer, M.G., Rassoulzadegan, F.: Are viruses driving microbial
  diversification and diversity?
\newblock Environ. Microbiol. \textbf{6}(1), 1--11 (2004)

\bibitem{wilmes2009dynamic}
Wilmes, P., Simmons, S.L., Denef, V.J., Banfield, J.F.: The dynamic genetic
  repertoire of microbial communities.
\newblock FEMS Microbiol. Rev. \textbf{33}(1), 109--132 (2009)

\bibitem{woese2002evolution}
Woese, C.R.: On the evolution of cells.
\newblock Proc. Natl. Acad. Sci. U.S.A. \textbf{99}(13), 8742--8748 (2002)

\bibitem{woese1977concept}
Woese, C.R., Fox, G.E.: The concept of cellular evolution.
\newblock J. Mol. Evol. \textbf{10}(1), 1--6 (1977)

\bibitem{woese1977phylogenetic}
Woese, C.R., Fox, G.E.: Phylogenetic structure of the prokaryotic domain: the
  primary kingdoms.
\newblock Proc. Natl. Acad. Sci. U.S.A. \textbf{74}(11), 5088 (1977)

\bibitem{woese2000aminoacyl}
Woese, C.R., Olsen, G.J., Ibba, M., Soll, D.: Aminoacyl-{tRNA} synthetases, the
  genetic code, and the evolutionary process.
\newblock Microbiol. Mol. Biol. Rev. \textbf{64}(1), 202 (2000)

\end{thebibliography}

\end{document}